\documentclass[usenatbib]{mn2e}

\usepackage{epsfig}
\usepackage{graphicx}
\usepackage{tabularx}
\usepackage{amsmath}
\usepackage{amssymb}
\usepackage{amstext}
\usepackage{amsfonts}

\newcommand{\cm}[1]{\unskip\ensuremath{\,{\rm cm}^{#1}}}

\newcommand{\ergs}[0]{\unskip\ensuremath{\,\textup{erg\,s}^{-1}}}
\newcommand{\bm}[1]{\ensuremath\mathbf{#1}}
\newcommand{\avrg}[1]{\ensuremath{\langle #1 \rangle}}

\newcommand{\pder}[2]{\ensuremath\frac{\partial #1}{\partial #2}}

\title{Morphology of flows and buoyant bubbles in the Virgo cluster}

\author[Georgi Pavlovski et al.]%
{Georgi Pavlovski$^1$%
\thanks{Email: gbp@phys.soton.ac.uk (GP)},
Christian R. Kaiser$^1$,
Edward C.D. Pope$^{1,2,3}$,
Hans Fangohr$^2$\\
$^1$ School of Physics and Astronomy, University of Southampton,
Southampton SO17 1BJ, U.K.\\
$^2$ School of Engineering Sciences, University of Southampton,
Southampton SO17 1BJ, U.K.\\
$^3$ School of Physics and Astronomy, University of Leeds, Leeds, 
LS2 9JT, U.K.
}
\date{Accepted .... . Received .... ; in original form .... }

\begin{document}

\maketitle

\begin{abstract}
There is growing evidence that the active galactic nuclei (AGN)
associated with the central elliptical galaxy in clusters of galaxies
are playing an important role in the evolution of the intracluster
medium (ICM) and clusters themselves.  We use high resolution
three-dimensional simulations to study the interaction of the cavities
created by AGN outflows (bubbles) with the ambient ICM.  The
gravitational potential of the cluster is modelled using the observed
temperature and density profiles of the Virgo cluster.  We demonstrate
the importance of the hydrodynamical Kutta-Zhukovsky forces associated
with the vortex ring structure of the bubbles, and discuss possible
effects of diffusive processes on their evolution.
\end{abstract}


\section{Introduction}
\label{int}
It has been more then 40 years since the first detection of x-ray
emission from a cluster of galaxies was made.  The emission was
detected from around the M87 galaxy \citep{Byram66} at the centre of
the Virgo cluster, whose proximity helped to get high resolution maps
of the x-ray emitting gas \citep[][hereafter G04]{Matsushita02,
Young02, Ghizzardi04}.  Later observations revealed that many clusters
are bright x-ray sources, with luminosities in the range
$10^{43\ldots45}$\ergs.  The properties of the x-ray emission from
clusters of galaxies were found to be most consistent with thermal
bremsstrahlung from hot gas \citep{Sarazin86}.  This implies that the
observed ICM has an electron density, $n_\text{e}$, in the range of
$10^{-4\ldots-2}\cm{-3}$, and a temperature, $T_\text{a}$, of the
order $10^{7\ldots8}$~K.

The radiative cooling time of the gas in a cluster core due to
bremsstrahlung can be as short as $10^6$~yr.  Unless the gas is
thermally supported, the cooling leads inexorably to an inflow of the
cold gas onto the central galaxy \citep[for a review of cooling flow
theory see, {\it e.g.},][]{Fabian94}.  However, many observations have since
demonstrated both the lack of the cold gas deposits, and
spectroscopically determined mass deposition rates up to an order of
magnitude smaller than the rates predicted by the classical cooling
flow model \citep{Voigt04}.

This apparent contradiction has led to the suggestion that the gas in
the cluster cores is reheated.  A number of different mechanisms were
proposed to reduce the cooling flow (or prevent it from forming in the
first place): currently the most popular candidates are outflows from
active galactic nuclei (AGN) \citep[see,
{\it e.g.},][]{Churazov01,Brueggen02,Birzan04}; thermal conduction of heat
from the outer regions \citep[see, {\it e.g.},][]{Narayan01,Voigt02,
Voigt04,Pope06}; disruption and heating of the flow by sound waves and
turbulence \citep[see, {\it e.g.},][]{Ruszkowski04,Fujita05, Fujita04b};
heating by supernovae explosions \citep[see, {\it e.g.},][]{Voit01,
Pipino02, Tang05}.  Clear discrimination between the different models
based on the observational data is difficult.  Morphological features
of the x-ray emission were used as the evidence in favour of a
particular mechanism \citep{Ruszkowski04,Fabian03,Brueggen05}.  The
morphological resemblance is important, and although it is impossible
to base a prove in favour of a particular mechanism on the morphology
alone, it is important to understand how different physical processes
affect the morphology of the observed features.

Despite some recent progress in the understanding of the physical
state of the ICM \citep{Ensslin05, Ensslin06, Lazarian06}, many key
parameters remain not well known.  Since laboratory experiments on
plasma with conditions close to those found in astronomical objects
are still rare \citep{Keenan04}, all estimates of such physical
properties of the ICM as the values of thermal diffusivity and
viscosity are generally based on modifications of the values
calculated in the classical work of \cite{Spitzer69}.  It is accepted
that magnetic fields decrease the diffusivity, since they suppress
movements of charged particles across the field lines.  The same is
believed to be true for the value of the viscosity as well.  Usually a
fraction of the Spitzer value is used to approximate thermal
diffusivity of the ICM \citep{Chandran98,Narayan01}, which assumes
chaotic orientation of the magnetic field lines ({\it i.e.}, magnetic
fields lines are bent on scales smaller than the mean free path of an
electron in the plasma).  The value of the thermal conductivity in
turbulent ICM can be an order of magnitude larger than the Spitzer
value, as shown by \cite{Cho03b} and \cite{Lazarian06}.  The exact
extent of this amplification depends on the state of the turbulence in
the ICM, which is difficult to estimate from the present observational
data.

The kinetic effects are likely to be important in the context of the
ICM physics.  Collisionless interactions on the scale smaller than the
mean free path can affect the overall dynamics of the
AGN-ICM interaction \citep{Schekochihin06, Schekochihin05}.

\subsection{The Model}
\label{mod} 
While a model that accounts for the above mentioned physical
properties of the ICM is clearly necessary for a detailed
understanding of its dynamics, we will use a simplified approach in
order to investigate certain aspects of the dynamics related to AGN
activity.  In this work we use a purely hydrodynamical model of the
ICM, which includes radiative cooling to simulate conditions
preexisting to AGN activity due to the cooling flow.

We use numerical simulations to study the morphology of the AGN-blown
bubbles.  We do not simulate inflation of the bubbles by AGN outflows,
instead we introduce them as perturbations of the temperature and
density fields of the cluster at a certain point during the simulation
(see \S\ref{ibub} for more details).  While the inflation of the
bubbles can influence their subsequent evolution, we are more
interested in the time scales that are larger than the inflation time,
and the effects of numerical techniques on the hydrodynamics of the
bubbles.  Although the effect of the velocity field produced by the jet
inflation of the bubbles is likely to be important especially during
the early stages of the evolution, here we study the influence of the
velocity field induced by the buoyantly rising bubble and the fluid
instabilities developing on its surface on the morphology of the
bubbles.  The combined effect of the jet injection and buoyancy will be
addressed in the future work.

The gravitational potential of the cluster is modelled using observed
temperature and density profiles for the Virgo cluster as determined
by G04, using XMM, BeppoSAX and Chandra data (see \S\ref{init} for
more details).  The observational data allows for more accurate
modelling of the gravitational potential then a fitted theoretical
profile \citep[{\it e.g.}, NFW,][]{NFW97}, as the gravity of the central
galaxy can become dominant at the cluster core.

The outline of the current work is as follows: in \S~\ref{num} we
discuss numerical framework and initial conditions for our
simulations, in \S~\ref{sres} we present analysis of the
simulations.  In \S~\ref{diss} we discuss implications and limitations
of the current work.  Finally, we present our main conclusions in
\S~\ref{conc}.
\section{Numerical Framework}
\label{num}
For our numerical experiment we use {\sc flash} (version 2.3) -- a
modular, adaptive-mesh (AMR), parallel simulation code, capable of
handling general compressible flow problems found in many
astrophysical environments \citep{Fryxell00}.

The {\sc hydro} module of the {\sc flash} code solves Euler's
equations in three dimensions.  In the conservative form the equations
are given by,
\begin{gather}
\label{eq:s1}
\pder{\rho}{t} + \nabla\cdot\left(\rho\bm{v}\right) = 0,\\
\label{eq:s2}
\pder{\rho\bm{v}}{t} + \nabla\cdot\left(\rho\bm{v}\otimes\bm{v}\right)
+ \nabla P = \rho \bm{g}\\
\label{eq:s3}
\pder{\rho {\cal E}}{t} + \nabla\cdot\left(\left(\rho {\cal E} +
P\right)\bm{v}\right) = \rho\left(\bm{v}\cdot\bm{g}\right)
\end{gather}
where $\rho$ is the fluid density, $\bm{v}$ is the fluid velocity, $P$
is the pressure, ${\cal E}$ is the sum of the internal energy $E$, and
the kinetic energy per unit mass, ${\cal E} = E + |\bm{v}|^2 / 2$,
$\Lambda$ is the cooling function (see \S~\ref{rcol}), and
$(\bm{v}\otimes\bm{v})_{ij}=v_iv_j$ denotes a tensor product.

The physical size of the computational grid is
$10^{24}\cm{}\approx322$~kpc in each direction.  Numerically the grid
is constructed using nested blocks of $16^3$ cells, with the ratio of
sizes between the neighbouring blocks being either 1:1 or 1:2.  The
minimum refinement level (minimum level of nested blocks with 1:2
ratio) is set to 3, which results in the largest cell size of
$10^{24}\cm{}/(2^{3-1}16)=10^{24}\cm{}/64\approx
1.56\times10^{22}\cm{}\approx5$~kpc.  The maximum refinement level was
set to 8, giving a minimum cell size of
$10^{24}\cm{}/(2^{8-1}16)=10^{24}\cm{}/2048\approx
4.88\times10^{20}\cm{}\approx0.16$~kpc.

\subsection{Radiative Cooling}
\label{rcol}
\begin{figure}
\centering\includegraphics[width=0.95\linewidth]{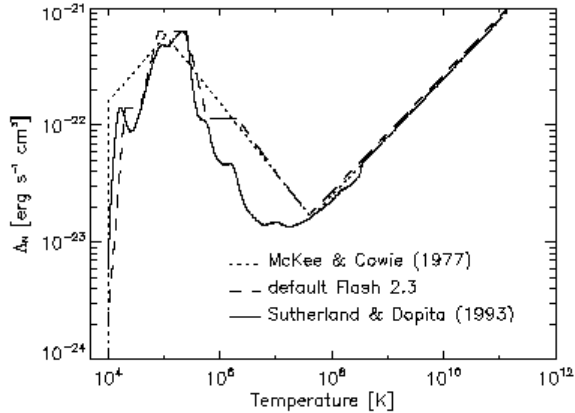}
\caption{Dependence of the cooling function, $\Lambda_\text{N}$, on
temperature (see \S~\ref{rcol}): thick line -- the current model;
dashed line -- the original {\sc flash} cooling function; dotted line
-- the cooling function from McKee \& Cowie (1977).}
\label{fig:lambda}
\end{figure}

We have developed a new module for {\sc flash}, which implements the
cooling function, $\Lambda$ (see Fig.~\ref{fig:lambda}), to account
for the radiative losses from the fully ionised plasma in the wide
range of temperatures, $4 < \log T < 8.5$, using the values tabulated
by \cite{Sutherland93} (metallicity was taken to equal half solar
[Fe/H]$= -0.5$).  For temperatures exceeding $\log T = 8.5$ we
calculate $\Lambda$ using the formula for thermal bremsstrahlung
\citep{McKee77},
\begin{gather}
\dot{E} = -\frac{1}{\rho}\Lambda,\\ \Lambda = n_\text{e} n_\text{p}
\Lambda_\text{N},\\ \Lambda_\text{N} =
2.5\times10^{-27}T^{0.5}\quad\text{[erg s$^{-1}$ cm$^3$]},
\end{gather}
where $\dot{E}$ is the rate of the energy loss per unit mass,
$n_\text{e}$, $n_\text{p}$ are the electron and proton number
densities, which for fully ionised gas with primordial abundances
(mass of Helium is 0.25 of total mass of the gas) relates to the mass
density of the ICM as,
\begin{gather}
n_\text{e} = 1.167n_\text{p},\\ \rho = 1.143 n_\text{e} a,
\end{gather}
where $a = 1.661\times10^{-24}$~g is the atomic mass unit ($\approx
m_\text{p}$).

The new cooling function accounts for line cooling in much greater
detail, but it does not differ significantly from the default {\sc
flash} module ({\sc source\_terms/cool/radloss}), see
Fig.~\ref{fig:lambda}.  The only serious departure from the default
approximation occurs around temperature of $T\sim10^7$~K, were the
original module overestimates the energy loss by a factor of a few
($\sim 5$).  It is important to note, that in our simulations most of
the gas has temperatures within this range, and therefore this
difference is non-negligible.  The radiative losses were not applied
to the gas with temperatures below $T=10^4$~K because the present
scheme does not thermally resolve gas at such temperatures (cooling
length becomes smaller than the numerical resolution of our grid).

\subsection{Initial Conditions}
\label{init}
\begin{figure*}
\centering\includegraphics[width=0.8\linewidth]{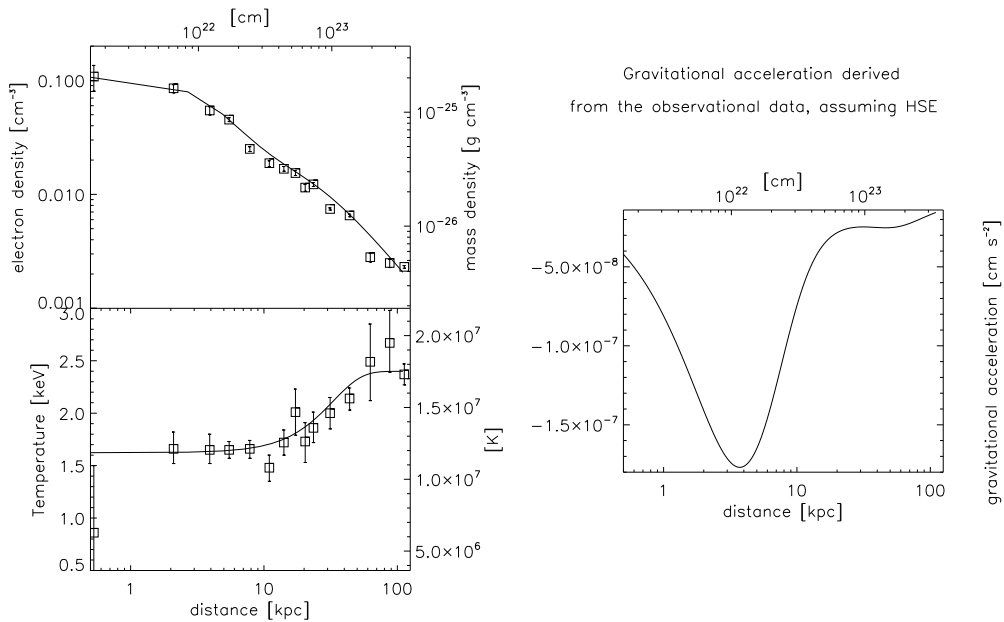}
\caption{Left panel: observed temperature and density profiles of the
Virgo cluster from G04. The lines are the analytical fit, see
equations (\ref{eq:prof1}) and (\ref{eq:prof2}). Right panel:
gravitational acceleration as a function of distance from the centre,
derived from HSE conditions using the best-fit lines.}
\label{fig:profiles}
\end{figure*}

We model the cluster gravitational potential using the data from G04.
Using the de-projected temperature and electron number density
profiles of the Virgo cluster,
\begin{gather}
\label{eq:prof1}
T(x) = T_1 - T_2\exp\{-x^2/(2x_2^2)\},\\ n_\text{e}(x) =
n_1\left(1+(x/x_1)^2\right)^{-a_1} +
n_2\left(1+(x/x_2)^2\right)^{-a_2},
\label{eq:prof2}
\end{gather}
where $x_1 = 1.54\times10^{22}\cm{}=5$~kpc, $x_2 =
7.19\times10^{22}\cm{}=23.3$~kpc, $a_1=1.518$, $a_2=0.705$,
$T_1=2.78\times10^7$~K, and $T_2=8.997\times10^6$~K,
$n_1=0.089\cm{-3}$, $n_2=0.019\cm{-3}$, (see Fig.~\ref{fig:profiles}),
we compute the gravitational acceleration due to the cluster's
gravitational potential from the assumption of hydrostatic equilibrium
(HSE),
\[
\frac{1}{\rho}\frac{{\rm d}P}{{\rm d}x} = g(x),
\]
where $P=n k_\text{B} T$ is the thermal pressure, and $g(x)$
is the gravitational acceleration.

At the beginning of the simulations we set the temperature field to be
uniform, $T_0 = 3\times10^7$~K, and the initial density distribution
is determined from the requirement of HSE.  Since HSE fixes only the
gradient of the density, we are left to choose the peak density,
$\rho_0$, {\it ad hoc}.  From a number of one dimensional test simulations
(without AGN heating) we determined the optimal value as $\rho_0=
5.15\times10^{26}$~g~cm$^{-3}$, which provides the best fit to the
observational profiles on the {\em outskirts} of the cluster at time
$t\sim 10^{10}$~yr after the beginning of the simulation \citep[see
also simulations by][]{Pope05}.

\subsection{Introduction of Bubbles}
\label{ibub}
We use a single fluid in {\sc flash} to represent both ambient ICM and
the gas inside bubbles, unlike the two-fluid approach used by
\cite{Brueggen05} and \cite{Ruszkowski04}.  The bubbles are introduced
into the simulation when the temperature of a computational cell in
the vicinity of the centre falls below $T=1.5$~keV.  This trigger
temperature serves as a simple feedback mechanism that relates cooling
flow parameters to the heating response.  The value $1.5$~keV was
selected to be close to the observed temperature minimum, see
Fig.~\ref{fig:profiles}.

The temperature field perturbations associated with the bubbles are
introduced using a modified version of the perturbation module {\sc
util/perturb}, which is supplied with {\sc flash}.  We position the
centres of the perturbations symmetrically on opposite sides of the
centre of the grid (cluster), at distances of
$x_0=1.5r_0=1.5\times10^{22}\cm{}=4.86$~kpc, with randomly specified
orientation angles of the line connecting the bubbles and the cluster
centre.  The locations for the bubbles are refined up to the maximum
refinement level prior to the introduction of the bubbles.

The temperature of all computational cells with coordinates $\bm{b}$,
satisfying the inequality $r_0\leq |\bm{b} - \bm{c}_i|$, where
$\bm{c}_i$ ($i=1,2$) are the position-vectors of the bubble centres,
is modified using the following algorithm.  Each computational cell
$\bm{b}$ is sub-divided into $4\times4\times4$ sub-cells ($\bm{b}_s$,
$s=1,\ldots,64$).  The temperature of each sub-cell is set to
$T_\text{b}=5\times10^{10}$~K if $r_0 < |\bm{b}_{s}-\bm{c}_i|$, and is
left unchanged otherwise.  The temperature of a computational cell is
then calculated as a volume average of temperatures of its sub-cells,
$T(\bm{b})= 1/64 \sum_s^{64} T(\bm{b}_{s})$.  This algorithm provided
a sharp transition boundary between the bubble and the ICM (so-called
top-hat perturbation profile), but is sufficiently robust not to cause
any numerical difficulties.

To make the temperature perturbations thermodynamically consistent, we
choose a pressure profile, and calculate the corresponding densities
using the equation of state for an ideal gas.  We tested two profiles:
the pressure inside the bubble matches the pressure of the ambient
medium ({\it i.e.} the pressure field is not changed by the perturbation),
and a constant pressure profile, where the value of the constant
pressure is calculated as a volume average of the pressure field
inside the perturbation region before the introduction of the bubbles.
However, we find that since the sound speed inside the bubble is much
higher than the sound speed of the ambient medium, in the former case
the pressure inside the bubble quickly reaches a constant profile,
sending a weak sound wave into the ICM.  In order to avoid this
additional artificial disturbance we used constant pressure profiles
in our setup.  Further analysis (see \S~\ref{sres}) showed that the
difference in the initial conditions does not lead to a serious
discrepancy in the properties of the bubbles.  The minor effects
caused by it will be discussed in \S~\ref{diss}.  As a first order
approximation, however, the difference between the parameters of the
bubbles in these two cases can be viewed as an intrinsic scatter
caused by our numerical scheme.

\section{Simulation Results}
\label{sres}
In this section we present an analysis of the data from simulations
with two kinds of the initial pressure profiles for bubbles, model
{\sc a} has the initial pressure profile matching the ambient pressure
and {\sc c} has the initial constant pressure profile.

The core of the cluster met the criterion for injection of bubbles
after $t=706$~Myr, so the bubbles are introduced into an established
cooling flow.  At this stage the ambient temperature of the ICM at 
the distance $x_0=1.5r_0$ from the centre is
$T(x_0)\approx2.62\times10^7$~K, density
$\rho(x_0)\approx2.83\times10^{-26}$~g~cm$^{-3}$, and the cooling time
$\Delta t\approx1.7$~Gyr.  Due to the large cooling time we can
ignore the effects of the radiative cooling on the parameters of the
bubbles.

In our purely hydrodynamical simulations we need to resolve the smallest
scales of the instabilities in order to account properly for the small
scale mixing of the hot gas with the ambient ICM.  The key parameter
in the stabilisation of the bubble/ambient medium interface against
the RT instability is the surface tension, which can be provided by
the magnetic field lines parallel to the surface of the bubble.  As it
was shown by \cite{Kaiser05} (hereafter K05) and \cite{DeYoung03} even
weak magnetic surface tension (magnetic fields of $\sim0.1\mu$G)
stabilise the bubble surface on scales up to 0.1~kpc at the distance
of $5$~kpc from the centre of the cluster (see Fig.~3, K05).  By
limiting the smallest cell size to $h=0.16$~kpc we essentially assume
that the gas will be fully mixed on the scale $\sim h$ on the time
scale of $h/c_\text{s}~\sim10^{13}$~s, which is approximately equal to
the growth time of the RT instability of the scale $h$ in the absence
of viscosity (see Fig.~2, K05).  Although a fully consistent model
should include magnetic fields to account for the stabilisation of the
bubble surface \citep{Ruszkowski07}, these simple estimates show that
the numerical resolution of our model is consistent with the expected
dynamics of the system, as follows from the analytical stability
analysis, and numerical simulations with `random' magnetic fields.

\subsection{Bubble Surface}
\label{bubs}

As the bubbles ascent in the ICM their shape and position changes.  In
order to study their parameters we have used a simple technique for
identification of the material inside bubbles at any stage during
their evolution.  Firstly, we note that the averaged radial entropy
index ($\sigma=\rho^{-2/3}T$) profile for the ICM, can be well fitted
with a linear function, $\avrg{\sigma} = \sigma_\text{a} x +
\sigma_\text{b}$, where $x$ is a distance from the centre, at any time
during the simulations in both models.  Secondly, by constructing the
entropy profile using the means over large enough spherical shells
({\it i.e.}, using low resolution profiling) we can abate the contribution
from the high entropy regions associated with the bubbles
($V_\text{bubble}/V_\text{shell}\ll 1$).  The variances of the values
of $\sigma$ in each shell are used as an error estimate for the values
of the means in the fitting routine, which further reduces influence
of the high entropy bubbles on the result of the linear fitting.  We
mark as ``bubble'' any computational cell with an entropy index above
$1.5(\sigma_\text{a} x + \sigma_\text{b})$, where $x$ is the distance
of the cell to the centre of the cluster (grid), and
$\sigma_{\text{a,b}}$ are the fitted coefficients for the
current time.

The surface of the bubbles becomes irregular due to RT instabilities,
and small bits of the bubble are shredded away as it rises through the
ICM and fragments.  In such circumstances a measure of the size and
position of the bubble material has to be essentially statistical, and
not just purely geometrical.

To quantify the size of the bubble we measure its extension in the
direction of ascent, $R_{||}$, and its size in the direction
perpendicular to the direction of ascent, $R_\bot$.  For a purely
spherical bubble $R_{||}=R_\bot$.  When the bubble has the shape of a
torus we measure both external and internal radii of the torus,
$R_{\bot(\text{max,min})}=\max,\min (R_\bot)$, using the following
algorithm.  The position vector of a ``bubble'' cell, $\bm{R}_i$, is
split into the component aligned with the direction specified by a
vector connecting the centre of the cluster and the initial position
of the bubble, $\bm{R}_{||i}$ (the direction of ascent), and the
vector perpendicular to the first one, going through the centre of the
cell, $\bm{R}_{\bot i}$, so that, $\bm{R}_i= \bm{R}_{||i}+
\bm{R}_{\bot i}$.  We then compute histograms (using 100 bins) of
$|\bm{R}_{\bot i}|$ and $|\bm{R}_{|| i}|$, using the values of the
corresponding vectors $\bm{R_{(||,\bot)i}}$ for all cells of a single
bubble.  We use the cut off at the level of 0.1 of the maxima of the
histogram to find the approximate extent of the bubble material (and
exclude any trailing and overtaking bits; we select the bins below the
cut-off which are closest to the peak of the histogram), that gives
the inner and outer radii of the torus, $R_{\bot\text{max,min}}$, or,
in the case of the 'quasi-spherical' bubble, two measures of the
radius, $R_{\bot}=R_{\bot\text{max}}$, and
$R_{||}=0.5(R_{||\text{max}}-R_{||\text{min}})$.  The means of the
$|R_{(||,\bot)i}|$ with the exclusion of the values below the cut-off,
correspond to the distance of the bubble from the centre of the
cluster (in the case of $|\bm{R}_{|| i}|$), and a mean radius of the
torus (in the case of $|\bm{R}_{\bot i}|$).

\subsection{Morphology}
\label{mrph}
The surfaces of bubbles rising in the ICM corrugates due to Reilegh-Taylor
(RT) instability, however, this does not result in an
immediate dispersion of the bubbles, see Fig.~\ref{fig:rtinst}.
In fact, the mass diffusion due to the RT instability and numerical
diffusion remains quite small and the bubbles in simulations {\sc a} and
{\sc c}  remain traceable for a long time.  We can
still find traces of the old bubble material after the
second pair of bubbles was injected ($>100$~Myr).

We find that during the ascent the size of the bubbles does not change
adiabatically according with the change of the ambient pressure with
distance from the centre of the cluster,
\begin{equation}
\label{e:volm}
V(x) = V_0 (P(x)/P_0)^{-1/\gamma},
\end{equation}
where $P(x)$ is the pressure profile, with $\gamma=5/3$.  Also the
velocity of the bubbles does not reach a terminal level, as it
is usually expected from a simple theoretical analysis \citep[see,
{\it e.g.},][]{Ensslin02}.

\begin{figure}
\centering\includegraphics[width=0.95\linewidth]{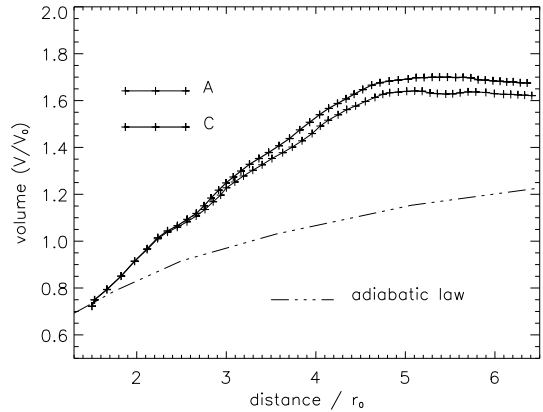}
\caption{Change of the volume of bubbles with height.  Note that due
to discretisation (finite resolution of the grid) and the algorithm we
use for identification of the bubble cells during the analysis, the
initial volume of the bubble is less than the geometrical value
$V_0=4/3\pi r_0^3$.  The dash-dotted line shows the predicted
isotropic adiabatic expansion of the spherical bubble according to
equation (\ref{e:volm}). The initial volume is taken to be equal to
the initial volume of bubble determined by the bubble finding
algorithm.}
\label{fig:volm}
\end{figure}

As shown in Fig.~\ref{fig:volm}, the volume of the bubbles keeps
growing after the injection and then it plateaus.  The growth of the
volume of the bubble (which is identified as the region of high
entropy) is the consequence of several different processes.  One of
them is the adiabatic expansion due to the pressure change with
distance from the centre of the cluster as given by equation
(\ref{e:volm}), another one is the entrainment of ambient plasma
through the surface of the bubble (mixing), and the third one is the
enlargement of the bubble due to the hydrodynamical Kutta-Zhukovsky
forces as explained below.  Full quantitative description of the
evolution of the bubble parameters is out of scope of this work, and
is addressed in the follow-up article \citep{Pavlovski06b}.

In our numerical experiment the entrainment of the ICM through the
boundary of the bubble is a consequence of the instabilities that
develop on the surface of the bubbles (see Fig.~\ref{fig:rtinst}),
and, unavoidably, the numerical diffusion.  In reality the diffusion
of the plasma across the boundary of the bubble is controlled by the
magnetic field, and the resulting change of the volume of the bubble
is likely to be different both from the prediction given by
equation~(\ref{e:volm}), and the results of our numerical model.  It
is important to note, however, that any amount of mixing will lead to
violation of the adiabatic assumption.  Unless the mixing process
is understood in greater detail, and other adiabatic processes like
Kutta-Zhukovsky forces are taken into account, the estimate of the AGN
energy input from the estimates of the total volume of the bubbles is
likely to be inaccurate.

\begin{figure}
\centering\includegraphics[width=0.95\linewidth]{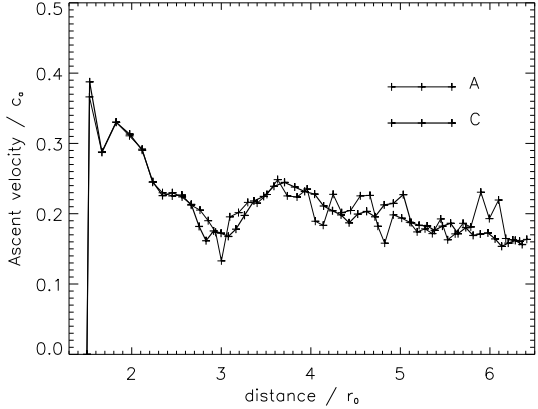}
\caption{Change of the ascent velocity of the bubbles with height.}
\label{fig:avel}
\end{figure}

The initial growth of the volume of the bubbles corresponds to the
deceleration phase in their ascent, which follows the very sharp
initial acceleration, see Fig.~\ref{fig:avel}.  The slight
re-acceleration of the bubbles at $x\approx 3r_0$ corresponds to the
change of the morphology of the bubbles from spherical to toroidal,
after which the bubbles continue to decelerate.  As explained below,
the velocity of the bubble is tightly linked with the increase in its
volume by a simple physical mechanism.

\begin{figure}
\centering\includegraphics[width=0.95\linewidth]{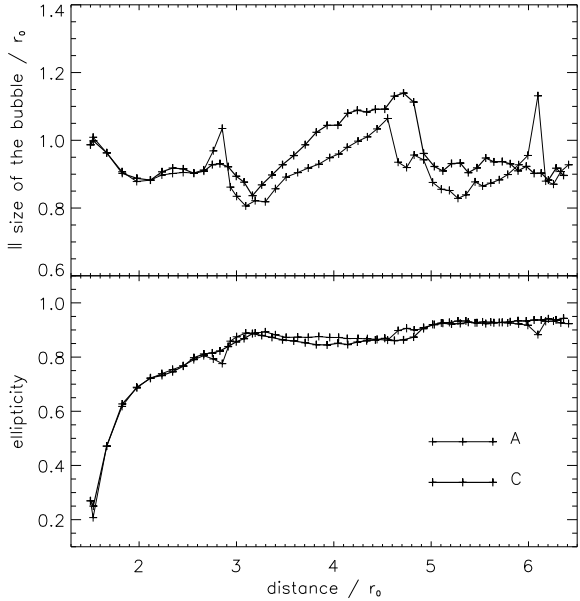}
\caption{Change of the size of the bubbles in the direction of the
ascent ($||$-direction), and the ellipticity, 
$\sqrt{1-(R_{||}/R_{\bot})^2}$, of the bubbles.}
\label{fig:brad}
\end{figure}

\begin{figure}
\centering\includegraphics[width=0.95\linewidth]{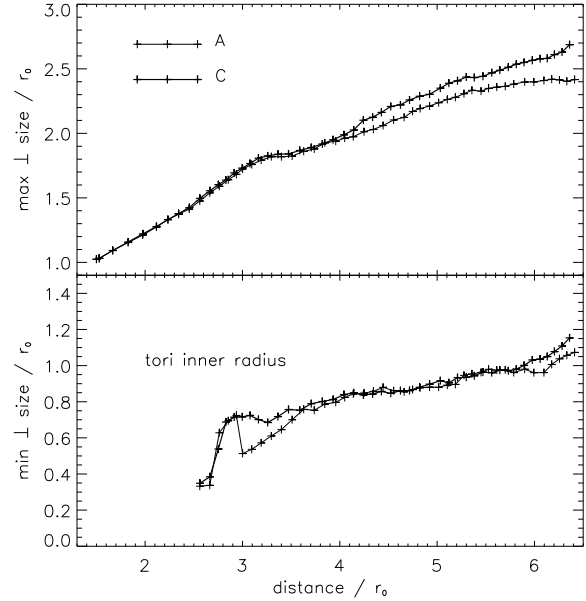}
\caption{Change of the size of the bubbles in the direction perpendicular
to the direction of ascent ($\bot$-direction).}
\label{fig:brad2}
\end{figure}

\begin{figure}
\centering\includegraphics[width=0.95\linewidth]{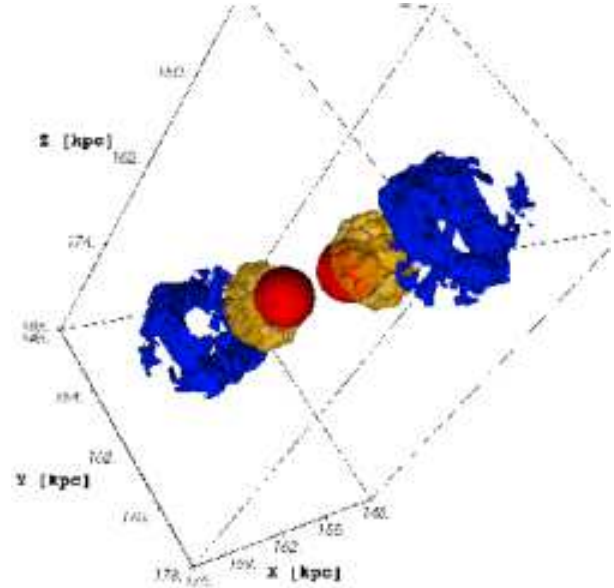}
\caption{Surfaces of the bubbles.  The bubbles closest to
the centre correspond to the time of the injection and are
spherically symmetric.  The later stages correspond to  
times 18~Myr and 98~Myr after the injection for (model {\sc c}). Note 
the `finger-like' corrugation of the surfaces of the bubbles.}
\label{fig:rtinst}
\end{figure}

The initial expansion of the bubble is not isotropic, see
Figs.~\ref{fig:brad} and \ref{fig:brad2}.  The volume of a bubble
increases due to the sideways expansion (expansion in the direction
perpendicular to the direction of ascent, $\bot$-direction), whereas
the size of the bubbles in the direction of the ascent
($||$-direction) stays roughly constant.  The ellipticity of the
bubbles, $e=\sqrt{1-(R_{||}/R_{\bot})^2}$, quickly reaches saturation
at a level of $e\approx0.9$ in both cases.

The flattening of the bubbles can not be attributed to the viscosity,
since no physical viscosity is modelled.  In presence of viscous
stresses they would act to squeeze the bubble in the $||$-direction.
They would result in an increase of pressure inside the
bubble, which would act to restore the equilibrium.  This process can
result in oscillations of the surface of the bubble (by analogy with
oscillations of the buoyant bubbles in a lava lamp), but is unlikely
to lead to any net increase in their volume.

\subsection{Kutta-Zhukovsky forces}
\label{ktzh}

\begin{figure}
\centering\includegraphics[width=0.95\linewidth]{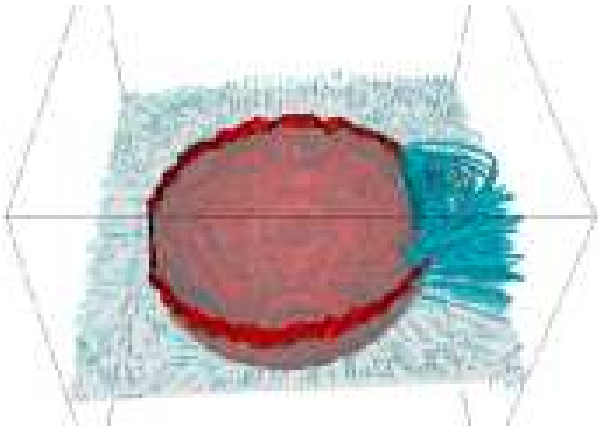}
\caption{The flow around the spherical bubble shows the structure
of a vortex ring.  The visualisation shows part of the
surface of the bubble (the upper half is cut off) 
in simulation {\sc c}, 2.3~Myr after the injection.
The arrows show the direction of the velocity in a thin slab just
below the cut plane.  The tubes trace the stream lines
close to the centre of the right vortex on the cut plane.}
\label{fig:flow}
\end{figure}

The structure of the flow through the bubble has the geometry of a vortex
ring right from the very early stages, when the bubble is still
roughly spherical, see Fig.~\ref{fig:flow}.  This circumstance
has a profound effect on the evolution of the bubble.

As a hot bubble starts to ascend, it can retain its spherical shape
only if the fluid motion inside the bubble correspond to a perfect
fluid dipole\footnote{There is a large amount of literature about
(quasi-) two-dimensional vortex dynamics, see, {\it e.g.},
\cite{Turner57,Morton60, Turner69,Fraenkel72}, see also
\cite{Afanasyev06} for a review of experiments.}  \citep[by analogy
with so-called Hill's solution, see, {\it e.g.},][]{Saffman95}.  If during
the ascent the spherical symmetry breaks down ({\it i.e.}, $R_\bot$ becomes
larger than $R_{||}$), the fluid motion inside the bubble corresponds
to the circulation around a stretched dipole -- the vortex ring.

\begin{figure}
\centering\includegraphics[width=0.7\linewidth]{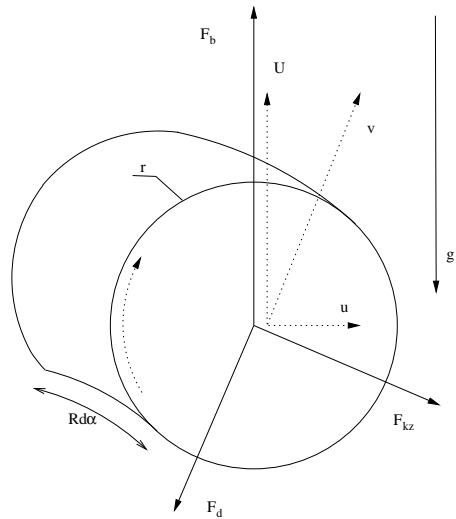}
\caption{The forces acting on an element of the vortex ring: $F_\text{b}$
- buoyancy, $F_\text{d}$ - drag, $F_\text{kz}$ - Kutta-Zhukovsky force.  Dashed
arrows show the components of the velocity: the ascent velocity $U=\dot{x}$
and the expansion velocity $u=\dot{R}$.}
\label{fig:vor}
\end{figure}

In the beginning we can consider the inner radius of the vortex
ring (torus) to be equal to zero (the bubble is roughly spherical), 
and the velocity of ascent to be equal to the maximum ascent 
velocity, see Fig.~\ref{fig:avel}.  Due to drag the 
speed of the ascent and of the vortex flow will decrease.  
Since the vortex moves with respect to the ambient medium each 
element of the vortex is subject to the 
Kutta-Zhukovsky\footnote{Zhukovsky's surname is sometimes 
spelt as Joukovsky or Joukowsky in the literature. See, {\it e.g.}, 
Joukowsky transform, also Kutta-Schukowski, Kutta-Joukowski, etc.}
force (for an introduction to aerodynamic forces see, {\it e.g.},
\cite{LandauIV} \S~38, and appendix~\ref{appendixb}), which  
acts to expand the vortex radially.  This force acts to 
expand the bubble in the $\bot$-direction, eventually creating
a hole in the middle of the bubble -- the bubble becomes torus
shaped.  As the vortex ring expands radially, 
an additional component of the Kutta-Zhukovsky force pushes
the bubble down, reducing its ascent velocity.  Indeed, with
rapid vortex expansion the total velocity vector of the elements
of the vortex ring will no longer be aligned in the $||$-direction, 
but will have a component in the $\bot$-direction as well, see 
Fig.~\ref{fig:vor}.  This results in the Kutta-Zhukovsky force pointing
down in the $||$-direction, in the direction opposite to the velocity 
of ascent.  This force will act to reduce the velocity of the ascent, 
{\it i.e.}, it counteracts buoyancy.

Let $R=\langle R_{||}\rangle$ be the radius of the torus, and $r$ the
radius of its cross-section, see Fig.~\ref{fig:vor}.  
If $\alpha$ is the central angle, so that $R\text{d}\alpha$ is 
the length of an element of the vortex ring, then this element will 
be subject to a Kutta-Zhukovsky force,
\begin{equation}
\label{eq:zhuk}
\Gamma U' \rho_\text{a}\left(1 + 
\left(\frac{\dot{R}}{U'}\right)^2\right)^{1/2}
R\text{d}\alpha,
\end{equation}
where $\dot{R} =\text{d}R/\text{d}t$ is the speed of the vortex
expansion, $\Gamma$ is the vortex circulation (see
appendix~\ref{appendixb}), $\rho_\text{a}$ is the density of the ICM,
and $U' = U - v'$, $U$ is the ascent velocity, $v'$ is the
self-induced (vertical) velocity of the vortex \citep{Batchelor67},
\begin{equation}
\label{eq:vind}
v'\approx\frac{\Gamma}{4\pi R}\log\frac{R}{r}.
\end{equation}
The Kutta-Zhukovsky force points at an angle $\beta$ to the $\bot$-direction,
\begin{equation}
\label{eq:zhukangle}
\begin{split}
\sin\beta=\frac{\dot{R}/U'}{\sqrt{1+\left(\dot{R}/U'\right)^2}},\\
\cos\beta=\frac{1}{\sqrt{1+\left(\dot{R}/U'\right)^2}}.\\
\end{split}
\end{equation}
So, the vertical component of the Kutta-Zhukovsky force, 
acting on the whole vortex ring is,
\begin{equation}
\label{eq:zhukpar}
F_{||\text{kz}}  = 2\pi R \Gamma \dot{R} \rho_\text{a}.
\end{equation}

\begin{figure}
\centering\includegraphics[width=0.95\linewidth]{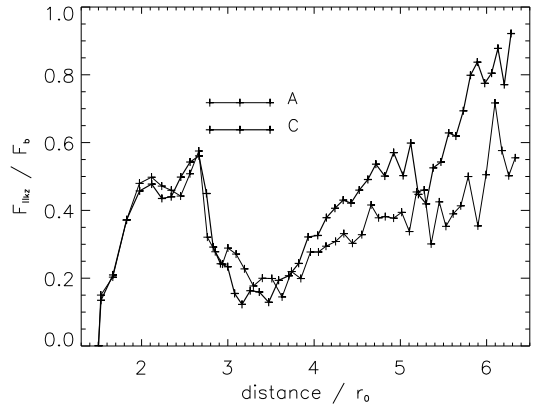}
\caption{Change of the ratio of the magnitude of the 
$||$-component of the Kutta-Zhukovsky force to the magnitude of 
the buoyant force.}
\label{fig:fkz}
\end{figure}

In Fig.~\ref{fig:fkz} we plot the ratio of the vertical component of
the Kutta-Zhukovsky force to the buoyant force acting on the bubbles
as a function of their distance from the cluster centre.  To estimate
the circulation of the vortex we calculated the mass-weighted mean of
the velocity of the bubble material in the rest frame of the bubble.
This is clearly just an estimate, and the resulting plot has some
scatter as a result.  It does, however, illustrate the main points
discussed above, and shows that the magnitude of the Kutta-Zhukovsky
force is comparable to the buoyant force.  A quantitative model of
the evolution of the bubbles is out of the scope of this work 
\citep[see][for more details]{Pavlovski06b}.

\section{Discussion}
\label{diss}
In the present work we have omitted the phase during which the jets
inflate the bubbles, which limits applicability of this model to the
interpretation of observations.  However, the dynamics of the flow
around bubbles created by a jet is likely to be similar, once the
bubbles become buoyant.  Taking the jet injection into account it is
quite possible that the vortex ring structure will form much earlier
(during the injection) and the flow around the bubble will have a
larger initial value of circulation, $\Gamma$, which can be induced by
the viscous momentum transfer of the kinetic energy of the jet, as
suggested by the laboratory experiments of small scale jets
\citep[see, {\it e.g.},][]{Afanasyev06}.  Kutta-Zhukovsky forces are then
likely to be even more important for the dynamics and the overall
energy balance.

The parameters of the bubbles with different initial pressure profiles
(models {\sc a} and {\sc c}) do not differ greatly, see
Figs.~\ref{fig:volm} -- \ref{fig:brad2}.  The only
noticeable difference is that the bubbles in simulation {\sc a}
develop a comparatively large secondary torus.  This secondary vortex
ring (torus) is shredded from the main vortex soon after the bubbles
morph into tori, it separates from the much bigger main torus, and
travels in front of it.  The volume of this secondary torus is small
compared to the volume of the bubble, and its contribution to the size
of the bubble was discarded (it has generally fallen below the 10 per
cent cut-off, see \S~\ref{bubs}) during the calculations.  The
secondary vortex ring quickly mixes with the ambient ICM and
dissipates.  The fact that this vortex ring has a larger volume in
case {\sc a} is reflected in the evolution of the total volume of the
bubbles, see Fig.~\ref{fig:volm}.  The volume of the {\sc a}-bubbles
is getting smaller than the volume of the {\sc c}-bubbles once their
secondary tori mix with the ICM.  It also results in a slightly
smaller $R_{\bot}$ size of the {\sc a}-bubbles at later stages of the
evolution, see Fig.~\ref{fig:brad2}.
\subsection{The role of missing physics}
\label{rthc}
High temperature plasma in the absence of the magnetic fields is
thought to be thermally conductive and viscous \citep{Spitzer69}. The
diffusivity coefficients are sensitive functions of temperature,
$\propto T^{2.5}$.  This ensures that any temperature gradient on 
scales larger than the mean free path of an electron is smoothed out very
efficiently by the conduction, and fluid instabilities are damped by
the viscosity \citep{Kaiser05}.

We have attempted to simulate the dynamics of bubbles in a viscous and
thermally conductive ICM.  Our experiments showed, however, that unless
thermal conduction is very efficiently suppressed on the boundary of the
bubble, it leads to a very fast erosion of the bubble material
(although no fluid instabilities develop).  This is easy to understand
from the following simple estimate.  Given that the amount of heat in the
bubble is of the order $\Delta Q\sim pV\approx10^{55}$~erg, the difference in
the temperatures of the plasma inside the bubble and the ambient ICM
$\Delta T\sim T\approx10^9$~K, the coefficient of the thermal
conductivity for plasma at $T=10^7$~K is
$\kappa\sim10^{11}$~erg~s$^{-1}$~cm$^{-1}$~K$^{-1}$, then the timescale of
the evaporation of the bubble is $\Delta t = \Delta Q \Delta l/(\kappa
\Delta T S)\sim1$~Myr, where we let $\Delta l\sim R$, and $S\sim R^2$.
Therefore it is not at all surprising that we found that the bubbles
were fully mixed with ICM at a distance of $\sim3\ldots4r_0$ from the
centre of the cluster.  Without the inclusion of magnetic fields into
the numerical model it is impossible to draw a definitive
conclusion about the evolution of the bubbles in a diffusive ICM.
Magnetic field draping \citep{Ruszkowski07} is likely to play a
significant role in the dispersion of the bubbles in thermally
conductive ICM.

We note, however, that the above discussion of the KZ forces does
still remain valid even in our `superdiffusive' cases (see
appendix~\ref{appendixa} for further details), {\it i.e.} the bubbles
still morph into tori, and the vortex motion is present.

The morphology of the observed bubbles in the Perseus cluster was used
as an indicator of a possible highly viscous ICM in \citep{Reynolds05}.
We would argue, however, that the flattening of the bubble, which was
explained by \cite{Reynolds05} as a result of viscosity, has an
additional explanation.  It is the Kutta-Zhukovsky forces, not
viscosity, that determine the radial size of the bubbles, and it is
impossible to infer the importance of viscous effects in the ICM from
the morphology of the bubbles alone.  Viscosity acts to suppress RT
instabilities but does not prevent morphing of the initially spherical
bubbles into tori.  At this stage the presence of a
significant viscosity in galaxy clusters should be regarded as
speculative, unless directly confirmed by observations.

\section{Conclusions}
\label{conc}

In the present work we have used a numerical scheme based on the {\sc
flash} code to produce three dimensional simulations of AGN bubble
heating of the ICM, with an emphasis on the dynamics of the bubbles.  Our
setup includes a sophisticated cooling function to help establish the
initial cooling flow, followed by the introduction of the high
contrast AGN-bubbles, which dynamics we have analysed.

The data presented in this article are consistent with the results of
previous numerical simulations of the AGN bubble heating of the ICM
\citep{Gardini06,Reynolds05,Brueggen02}.  However, our analysis has
indicated additional important phenomena linked to the dynamics of the
gas of the bubbles in the ICM.  Our main results are as follows.

\begin{enumerate}

\item Bubbles change shape and transform into tori not because
of hydrodynamical instabilities, but due to the radial Kutta-Zhukovsky
force.

\item Bubbles do not reach a terminal velocity defined by the balance
between the drag force and the buoyancy. The velocity of their ascent
is also affected by the the vertical Kutta-Zhukovsky force and
the self-induced velocity of the vortex.

\item Viscosity does play a role in stabilising the surface of the
bubbles against instabilities, but it does not alter their
overall 3D (torus-like) morphology.

\item The volume of bubbles can lead to an erroneous estimation of the
energy supplied by the AGN when using equation (\ref{e:volm}). The
expansion of the bubbles is not only due the change of the ambient
pressure but also due to the Kutta-Zhukovsky forces, and non-adiabatic
turbulent mixing.  A more accurate model of plasma (including magnetic
fields) is needed in order to make realistic prediction about the
entrainment of the ICM into the bubbles.

\item Thermal conduction must be highly suppressed on the boundary of
  the bubbles, otherwise it leads to a rapid erosion (mixing) of the
  bubble material.

\end{enumerate}

A better knowledge of the physical state of the ICM is of paramount
importance for further progress in understanding the interaction of
AGN outflows with the ICM.

\section{Acknowledgements} 
This research has made use of NASA's Astrophysics Data System
Bibliographic Services.  The software used in this work was in part
developed by the DOE-supported ASC / Alliance Center for Astrophysical
Thermonuclear flashes at the University of Chicago.  We would like to
thank {\sc flash} developers Tomasz Plewa and Timur Linde for prompt
response to our questions.  GP and CRK thank PPARC for financial
support.

\appendix
\section{Nature of the Kutta-Zhukovsky forces} 
\label{appendixb}
Circulation is the line integral around a closed curve of the fluid
velocity. If $\bm{v}$ is the fluid velocity and d$\bm{s}$ is a unit
vector along the closed curve $C$, the circulation is given by,
\[
    \Gamma=\oint_{C}\bm{v}\cdot\text{d}\bm{s}.
\]
The units of circulation are [cm$^2$~s$^{-1}$].  The {\em
Kutta-Zhukovsky theorem} states that the lift force acting per unit
span on a body in an inviscid flow field can be expressed as the
product of the circulation about the body, $\Gamma$, the fluid
density, $\rho$, and the speed of the body relative to the
free-stream, $U$,
\[
    f = U\rho\Gamma.
\]
This equation applies both around airfoils, where the circulation is
generated by airfoil action, and around spinning objects, experiencing
the Magnus effect, where the circulation is induced mechanically.  The
latter case is governed by the same basic principles as apply to the
travelling vortices.  A spinning object ({\it e.g.}, a cylinder rotating
around its axis of symmetry, or an eddy) creates a motion of the fluid
in the boundary layer around it ({\it e.g.}, via viscous drag) -- the
circulation.  If such a rotating object travels through the fluid
({\it e.g.},
the rotating cylinder travels through the fluid at rest in the laboratory
frame of reference, with its axis of symmetry remaining parallel to
itself), on one side of it, the velocity of the fluid in the boundary
layer will be in the same direction as the velocity of the surrounding
fluid near it.  On this side the resulting velocity of the fluid will
be larger than the travel velocity.  On the other side of the object,
the velocity of the fluid in the boundary layer will be in the
opposite direction of the velocity of the fluid near it, and the
resulting velocity of the fluid will be smaller than the travel
velocity.  According to the Bernoulli's theorem the quantity
$p+v^2/2=$~const along flow lines, so the pressure $p$ will be
lower on one side of the rotating cylinder than on the other causing
an unbalanced force at a {\bf right angle} to the travel velocity.

\section{Numerical experiments with thermal conductivity}
\label{appendixa}

In order to account for the suppression of the diffusive processes in
the presence of chaotic magnetic fields the diffusive coefficients for
conductivity, $\kappa_\text{s}$, and viscosity, $\eta_\text{s}$, are
reduced by a certain factor, $0\leq f_\text{s}\leq 1$.  In the case of
AGN blown bubbles this suppression is likely to be much stronger
inside the bubbles (due to the associated strong magnetic field),
resulting in the suppression coefficient being not spatially uniform,
\begin{gather}
\kappa = f_\text{s}(\bm{x}) \kappa_\text{s},\\ \eta =
f_\text{s}(\bm{x}) \eta_s,\\ f_\text{s}(\bm{x})=\begin{cases}
f_{\text{s}1}, & \text{outside bubbles},\\ f_{\text{s}2}, &
\text{inside bubbles}.
       \end{cases}
\end{gather}
The factor $f_{\text{s}1}$ is a simulation parameter\footnote{Note
that the suppression factors for viscosity and thermal conductivity
are likely to be similar, despite the fact that the corresponding
Larmor radii are different by several orders of magnitude.  In the
tangled magnetic field with the coherence scale $l_\text{B}$ the
Rechester-Rozenbluth distance \citep[for details see][]{Narayan01} for
electrons is given by, $L_\text{RR}\sim
l_\text{B}\ln(l_\text{B}/r_\text{e})\sim 10 l_\text{B}$, where
$r_\text{e}$ is Larmor radius for electrons.  For ions this distance
is very similar $L_\text{RR}\sim 25 l_\text{B}$, implying similar
suppression coefficient for both transport processes.}.  We have
tested a range of possible values: $f_{\text{s}1}=[0.1,0.3,0.6]$.  In
order to identify the computational cells that are located inside the
bubbles and to apply a different suppression factor to them, we probe
cells inside the spherical central region of the size $x_7 = 7r_0 =
7\times10^{22}\cm{}=22.7$~kpc, where $r_0$ is the initial radius of
the bubbles, and $x_7=7 r_0$ is an estimate of the maximum height the
bubbles are likely to reach, starting at the distance of $1.5r_0$ from
the centre of the cluster (see also \S~\ref{ibub}).  At each time-step
we compare the entropy index, $\sigma = T \rho^{-2/3}$, of the cells
inside the $|\bm{x}|<x_7$ region with the cutoff value, $\sigma_7$.
The cutoff value is computed for each time-step as,
$\sigma_7=1.5\langle\sigma\rangle_{x_7<|\bm{x}|\le x_7+\Delta x}$,
which is 1.5 times the average of the entropy index in a spherical
shell with radius $x_7$, and thickness $\Delta x=0.5r_0 =
5\times10^{21}\cm{}=1.6$~kpc.  We then define $f_{s2}$ as,
\begin{equation}
f_{\text{s}2} =
\begin{cases}
5.\times10^{-12} \left(\frac{2 \sigma_7}{3 \sigma}\right)^{1.5}, &
\text{if}\, \sigma > \sigma_7,\\ f_{\text{s}1}, &\text{otherwise},
\end{cases}
\end{equation}
where the factor $\left(2\sigma_7/3\sigma\right)^{1.5}\approx
(T_\text{a}/T_\text{b})^{2.5}$ accounts for the increase of the
conductivity and viscosity coefficients due to the large temperature
of the bubbles, $T_\text{b}$, compared to the temperature of the
ambient gas, $T_\text{a}$.  This scheme is designed to effectively
suppress diffusive process inside the bubbles.  It could not, however,
prevent diffusion of the bubble material and its mixing with the ICM.

\begin{figure}
\centering\includegraphics[width=0.95\linewidth]{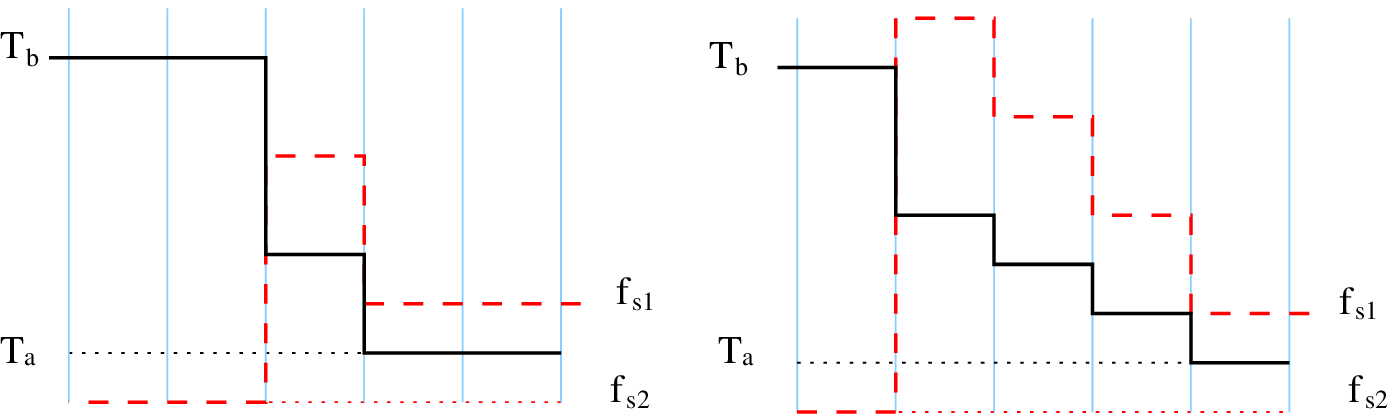}
\caption{Schematic of the growth of a conductive layer around the
bubble due to the discretization of the temperature gradient.  The
vertical lines represent boundaries of the computational cells.  The
solid line represents the temperature profile at the edge of the
bubble, where $T_\text{b}$ is the temperature of the gas inside the
bubble, and $T_\text{a}$ is the temperature of the ambient medium.
The dashed line represents the corresponding value of the thermal
conduction.  The suppression coefficient $f_{\text{s}1}$ is applied to
the gas at temperature $T_\text{a}$, and $f_{\text{s}2}$ to the gas at
temperature $T_\text{b}$ .  The panel on the left represents the
initial profiles.  Thermal conduction works to reduce the temperature
gradient, and creates a number of cells with an intermediate values of
temperatures (entropies) and high conductivities (right panel).}
\label{fig:mcond}
\end{figure}

\begin{figure}
\centering\includegraphics[width=0.95\linewidth]{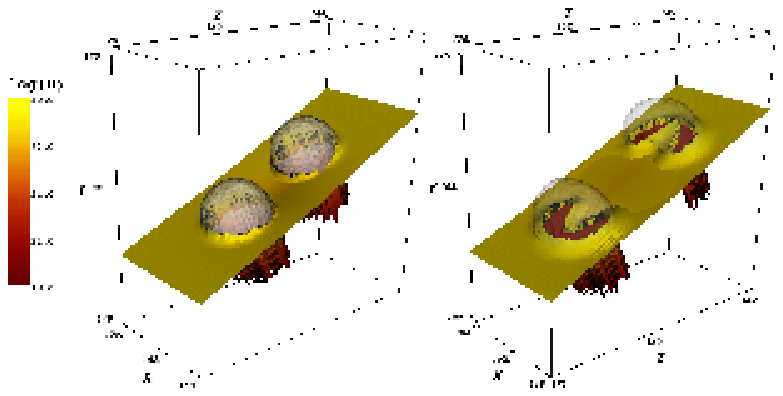}
\caption{Growth of a conductive layer around the bubbles in a 3D
simulation.  The visualisation shows surface cuts through the scalar
field equal to the logarithm of the thermal diffusivity.  The
resulting surfaces are colour-coded and warped according to the value
of the scalar.  Semi-transparent surfaces show the edge of the
bubbles.  The data are taken from the simulation with
$f_\text{s1}=0.1$: on the moment of injection (left), 8~Myr later
(right).  The size of the box is given in kiloparsecs.}
\label{fig:logd}
\end{figure}

In our diffusive simulations discretization of the temperature gradient
leads to formation of a layer of thermally conductive plasma around
the bubble, as illustrated in Figs.~\ref{fig:mcond} and
\ref{fig:logd}.  This layer is essentially an artifact of our
numerical scheme for the suppression of the diffusive processes, and is
not reflective of the real physical properties of the ICM.

The problem with discretization is going to be present in the
two-fluid approach as well (by two-fluid approach we mean separate
fluids for the bubble and the ambient medium as used by, {\it e.g.},
\cite{Brueggen05}).  We believe it will lead to a qualitatively
similar behaviour, since the suppression coefficient for the diffusion
in this case is also based on an {\it ad hoc} cut off based on the
concentration of the bubble fluid in a computational cell.

Without a more advanced model, which needs to include magnetic fields,
and self-consistently accounts for the suppression of the diffusive
processes, it is impossible to say whether or not this is a realistic
representation of the real state of the ICM.

We find that the dynamics and morphology of the bubbles in our 
diffusive simulations also governed by KZ forces, since the
circulative flow around the bubble is always present.

\bibliography{gbp}

\begin{thebibliography}{}

\bibitem[\protect\citeauthoryear{{Afanasyev}}{{Afanasyev}}{2006}]{Afanasyev06}
{Afanasyev} Y.~D.,  2006, Physics of Fluids, 18, 037103

\bibitem[\protect\citeauthoryear{{Batchelor}}{{Batchelor}}{1967}]{Batchelor67}
{Batchelor} G.~K.,  1967, An introduction to fluid dynamics.
Cambridge University Press

\bibitem[\protect\citeauthoryear{{B{\^i}rzan}, {Rafferty}, {McNamara}, {Wise}
  \& {Nulsen}}{{B{\^i}rzan} et~al.}{2004}]{Birzan04}
{B{\^i}rzan} L.,  {Rafferty} D.~A.,  {McNamara} B.~R.,  {Wise} M.~W.,
  {Nulsen} P.~E.~J.,  2004, \apj, 607, 800

\bibitem[\protect\citeauthoryear{{Br{\"u}ggen} \& {Kaiser}}{{Br{\"u}ggen} \&
  {Kaiser}}{2002}]{Brueggen02}
{Br{\"u}ggen} M.,  {Kaiser} C.~R.,  2002, Nat., 418, 301

\bibitem[\protect\citeauthoryear{{Br{\"u}ggen}, {Ruszkowski} \&
  {Hallman}}{{Br{\"u}ggen} et~al.}{2005}]{Brueggen05}
{Br{\"u}ggen} M.,  {Ruszkowski} M.,    {Hallman} E.,  2005, \apj, 630, 740

\bibitem[\protect\citeauthoryear{{Byram}, {Chubb} \& {Friedman}}{{Byram}
  et~al.}{1966}]{Byram66}
{Byram} E.~T.,  {Chubb} T.~A.,    {Friedman} H.,  1966, Science, 152, 66

\bibitem[\protect\citeauthoryear{{Chandran} \& {Cowley}}{{Chandran} \&
  {Cowley}}{1998}]{Chandran98}
{Chandran} B.~D.~G.,  {Cowley} S.~C.,  1998, Physical Review Letters, 80, 3077

\bibitem[\protect\citeauthoryear{{Cho}, {Lazarian}, {Honein}, {Knaepen},
  {Kassinos} \& {Moin}}{{Cho} et~al.}{2003}]{Cho03b}
{Cho} J.,  {Lazarian} A.,  {Honein} A.,  {Knaepen} B.,  {Kassinos} S.,
  {Moin} P.,  2003, \apjl, 589, L77

\bibitem[\protect\citeauthoryear{{Churazov}, {Br{\"u}ggen}, {Kaiser},
  {B{\"o}hringer} \& {Forman}}{{Churazov} et~al.}{2001}]{Churazov01}
{Churazov} E.,  {Br{\"u}ggen} M.,  {Kaiser} C.~R.,  {B{\"o}hringer} H.,
  {Forman} W.,  2001, \apj, 554, 261

\bibitem[\protect\citeauthoryear{{De Young}}{{De Young}}{2003}]{DeYoung03}
{De Young} D.~S.,  2003, \mnras, 343, 719

\bibitem[\protect\citeauthoryear{{En{\ss}lin} \& {Heinz}}{{En{\ss}lin} \&
  {Heinz}}{2002}]{Ensslin02}
{En{\ss}lin} T.~A.,  {Heinz} S.,  2002, \aap, 384, L27

\bibitem[\protect\citeauthoryear{{En{\ss}lin} \& {Vogt}}{{En{\ss}lin} \&
  {Vogt}}{2006}]{Ensslin06}
{En{\ss}lin} T.~A.,  {Vogt} C.,  2006, \aap, 453, 447

\bibitem[\protect\citeauthoryear{{En{\ss}lin}, {Vogt} \& C.{
  Pfrommer}}{{En{\ss}lin} et~al.}{2005}]{Ensslin05}
{En{\ss}lin} T.~A.,  {Vogt} C.,    C.{ Pfrommer} 2005, in 'The Magnetized
  Plasma in Galaxy Evolution' Krakow, Poland, Sept. 27th - Oct. 1st, 2004
  {Magnetic Fields in Clusters of Galaxies}.
p.~231

\bibitem[\protect\citeauthoryear{{Fabian}}{{Fabian}}{1994}]{Fabian94}
{Fabian} A.~C.,  1994, \araa, 32, 277

\bibitem[\protect\citeauthoryear{{Fabian}, {Sanders}, {Crawford}, {Conselice},
  {Gallagher} \& {Wyse}}{{Fabian} et~al.}{2003}]{Fabian03}
{Fabian} A.~C.,  {Sanders} J.~S.,  {Crawford} C.~S.,  {Conselice} C.~J.,
  {Gallagher} J.~S.,    {Wyse} R.~F.~G.,  2003, \mnras, 344, L48

\bibitem[\protect\citeauthoryear{{Fraenkel}}{{Fraenkel}}{1972}]{Fraenkel72}
{Fraenkel} L.~E.,  1972, J. Fluid Mech., 51, 119

\bibitem[\protect\citeauthoryear{{Fryxell}, {Olson}, {Ricker}, {Timmes},
  {Zingale}, {Lamb}, {MacNeice}, {Rosner}, {Truran} \& {Tufo}}{{Fryxell}
  et~al.}{2000}]{Fryxell00}
{Fryxell} B.,  {Olson} K.,  {Ricker} P.,  {Timmes} F.~X.,  {Zingale} M.,
  {Lamb} D.~Q.,  {MacNeice} P.,  {Rosner} R.,  {Truran} J.~W.,    {Tufo} H.,
  2000, \apjs, 131, 273

\bibitem[\protect\citeauthoryear{{Fujita}, {Matsumoto} \& {Wada}}{{Fujita}
  et~al.}{2004}]{Fujita04b}
{Fujita} Y.,  {Matsumoto} T.,    {Wada} K.,  2004, Journal of Korean
  Astronomical Society, 37, 571

\bibitem[\protect\citeauthoryear{{Fujita} \& {Suzuki}}{{Fujita} \&
  {Suzuki}}{2005}]{Fujita05}
{Fujita} Y.,  {Suzuki} T.~K.,  2005, \apjl, 630, L1

\bibitem[\protect\citeauthoryear{{Gardini}}{{Gardini}}{2007}]{Gardini06}
{Gardini} A.,  2007, \aap, 464, 143

\bibitem[\protect\citeauthoryear{{Ghizzardi}, {Molendi}, {Pizzolato} \& {De
  Grandi}}{{Ghizzardi} et~al.}{2004}]{Ghizzardi04}
{Ghizzardi} S.,  {Molendi} S.,  {Pizzolato} F.,    {De Grandi} S.,  2004, \apj,
  609, 638

\bibitem[\protect\citeauthoryear{{Kaiser}, {Pavlovski}, {Pope} \&
  {Fangohr}}{{Kaiser} et~al.}{2005}]{Kaiser05}
{Kaiser} C.~R.,  {Pavlovski} G.,  {Pope} E.~C.~D.,    {Fangohr} H.,  2005,
  \mnras, 359, 493

\bibitem[\protect\citeauthoryear{{Keenan} \& {Rose}}{{Keenan} \&
  {Rose}}{2004}]{Keenan04}
{Keenan} F.~P.,  {Rose} S.~J.,  2004, Astronomy and Geophysics, 45, 18

\bibitem[\protect\citeauthoryear{Landau \& Lifshitz}{Landau \&
  Lifshitz}{1987}]{LandauIV}
Landau L.~D.,  Lifshitz E.~M.,  1987, Fluid Mechanics, second edn.
Vol.~4, Pergamon Press

\bibitem[\protect\citeauthoryear{{Lazarian}}{{Lazarian}}{2006}]{Lazarian06}
{Lazarian} A.,  2006, \apjl, 645, L25

\bibitem[\protect\citeauthoryear{{Matsushita}, {Belsole}, {Finoguenov} \&
  {B{\"o}hringer}}{{Matsushita} et~al.}{2002}]{Matsushita02}
{Matsushita} K.,  {Belsole} E.,  {Finoguenov} A.,    {B{\"o}hringer} H.,  2002,
  \aap, 386, 77

\bibitem[\protect\citeauthoryear{{McKee} \& {Cowie}}{{McKee} \&
  {Cowie}}{1977}]{McKee77}
{McKee} C.~F.,  {Cowie} L.~L.,  1977, ApJ, 215, 213

\bibitem[\protect\citeauthoryear{{Morton}}{{Morton}}{1960}]{Morton60}
{Morton} B.~R.,  1960, J. Fluid Mech., 9, 107

\bibitem[\protect\citeauthoryear{{Narayan} \& {Medvedev}}{{Narayan} \&
  {Medvedev}}{2001}]{Narayan01}
{Narayan} R.,  {Medvedev} M.~V.,  2001, \apjl, 562, L129

\bibitem[\protect\citeauthoryear{{Navarro}, {Frenk} \& {White}}{{Navarro}
  et~al.}{1997}]{NFW97}
{Navarro} J.~F.,  {Frenk} C.~S.,    {White} S.~D.~M.,  1997, \apj, 490, 493

\bibitem[\protect\citeauthoryear{{Pavlovski}, {Kaiser} \& {Pope}}{{Pavlovski}
  et~al.}{2007}]{Pavlovski06b}
{Pavlovski} G.,  {Kaiser} C.~R.,    {Pope} E.~C.~D.,  2007, {Dynamics of
  buoyant bubbles in clusters of galaxies}, submitted, arXiv:0709.1796

\bibitem[\protect\citeauthoryear{{Pipino}, {Matteucci}, {Borgani} \&
  {Biviano}}{{Pipino} et~al.}{2002}]{Pipino02}
{Pipino} A.,  {Matteucci} F.,  {Borgani} S.,    {Biviano} A.,  2002, New
  Astronomy, 7, 227

\bibitem[\protect\citeauthoryear{{Pope}, {Pavlovski}, {Kaiser} \&
  {Fangohr}}{{Pope} et~al.}{2005}]{Pope05}
{Pope} E.~C.~D.,  {Pavlovski} G.,  {Kaiser} C.~R.,    {Fangohr} H.,  2005,
  \mnras, 364, 13

\bibitem[\protect\citeauthoryear{{Pope}, {Pavlovski}, {Kaiser} \&
  {Fangohr}}{{Pope} et~al.}{2006}]{Pope06}
{Pope} E.~C.~D.,  {Pavlovski} G.,  {Kaiser} C.~R.,    {Fangohr} H.,  2006,
  \mnras, 367, 1121

\bibitem[\protect\citeauthoryear{{Reynolds}, {McKernan}, {Fabian}, {Stone} \&
  {Vernaleo}}{{Reynolds} et~al.}{2005}]{Reynolds05}
{Reynolds} C.~S.,  {McKernan} B.,  {Fabian} A.~C.,  {Stone} J.~M.,
  {Vernaleo} J.~C.,  2005, \mnras, 357, 242

\bibitem[\protect\citeauthoryear{{Ruszkowski}, {Br{\"u}ggen} \&
  {Begelman}}{{Ruszkowski} et~al.}{2004}]{Ruszkowski04}
{Ruszkowski} M.,  {Br{\"u}ggen} M.,    {Begelman} M.~C.,  2004, \apj, 615, 675

\bibitem[\protect\citeauthoryear{{Ruszkowski}, {En{\ss}lin}, {Br{\"u}ggen},
  {Heinz} \& {Pfrommer}}{{Ruszkowski} et~al.}{2007}]{Ruszkowski07}
{Ruszkowski} M.,  {En{\ss}lin} T.~A.,  {Br{\"u}ggen} M.,  {Heinz} S.,
  {Pfrommer} C.,  2007, \mnras, pp 400--+

\bibitem[\protect\citeauthoryear{{Saffman}}{{Saffman}}{1995}]{Saffman95}
{Saffman} P.~G.,  1995, Vortex Dynamics, second edn.
Cambridge University Press

\bibitem[\protect\citeauthoryear{{Sarazin}}{{Sarazin}}{1986}]{Sarazin86}
{Sarazin} C.~L.,  1986, Reviews of Modern Physics, 58, 1

\bibitem[\protect\citeauthoryear{{Schekochihin}, {Cowley} \&
  {Dorland}}{{Schekochihin} et~al.}{2006}]{Schekochihin06}
{Schekochihin} A.~A.,  {Cowley} S.~C.,    {Dorland} W.,  2006, ArXiv
  Astrophysics e-prints

\bibitem[\protect\citeauthoryear{{Schekochihin}, {Cowley}, {Kulsrud}, {Hammett}
  \& {Sharma}}{{Schekochihin} et~al.}{2005}]{Schekochihin05}
{Schekochihin} A.~A.,  {Cowley} S.~C.,  {Kulsrud} R.~M.,  {Hammett} G.~W.,
  {Sharma} P.,  2005, in 'The Magnetized Plasma in Galaxy Evolution' Krakow,
  Poland, Sept. 27th - Oct. 1st, 2004 {Magnetised plasma turbulence in clusters
  of galaxies}.
p.~200

\bibitem[\protect\citeauthoryear{{Spitzer}}{{Spitzer}}{1962}]{Spitzer69}
{Spitzer} L.,  1962, Physics of Fully Ionized Gases.
Inerscience, New York

\bibitem[\protect\citeauthoryear{{Sutherland} \& {Dopita}}{{Sutherland} \&
  {Dopita}}{1993}]{Sutherland93}
{Sutherland} R.~S.,  {Dopita} M.~A.,  1993, ApJS, 88, 253

\bibitem[\protect\citeauthoryear{{Tang} \& {Wang}}{{Tang} \&
  {Wang}}{2005}]{Tang05}
{Tang} S.,  {Wang} Q.~D.,  2005, \apj, 628, 205

\bibitem[\protect\citeauthoryear{{Turner}}{{Turner}}{1957}]{Turner57}
{Turner} J.~S.,  1957, Proc. Roy. Soc. A., 239, 61

\bibitem[\protect\citeauthoryear{{Turner}}{{Turner}}{1969}]{Turner69}
{Turner} J.~S.,  1969, Annu. Rev. Fluid Mech., 1, 29

\bibitem[\protect\citeauthoryear{{Voigt} \& {Fabian}}{{Voigt} \&
  {Fabian}}{2004}]{Voigt04}
{Voigt} L.~M.,  {Fabian} A.~C.,  2004, \mnras, 347, 1130

\bibitem[\protect\citeauthoryear{{Voigt}, {Schmidt}, {Fabian}, {Allen} \&
  {Johnstone}}{{Voigt} et~al.}{2002}]{Voigt02}
{Voigt} L.~M.,  {Schmidt} R.~W.,  {Fabian} A.~C.,  {Allen} S.~W.,
  {Johnstone} R.~M.,  2002, \mnras, 335, L7

\bibitem[\protect\citeauthoryear{{Voit} \& {Bryan}}{{Voit} \&
  {Bryan}}{2001}]{Voit01}
{Voit} G.~M.,  {Bryan} G.~L.,  2001, Nat., 414, 425

\bibitem[\protect\citeauthoryear{{Young}, {Wilson} \& {Mundell}}{{Young}
  et~al.}{2002}]{Young02}
{Young} A.~J.,  {Wilson} A.~S.,    {Mundell} C.~G.,  2002, \apj, 579, 560

\end{thebibliography}

\end{document}